\PassOptionsToPackage{prologue,dvipsnames,table,xcdraw}{xcolor}
\documentclass[10pt,twocolumn,letterpaper]{article}

\usepackage[pagenumbers]{cvpr} 

\definecolor{cvprblue}{rgb}{0.21,0.49,0.74}
\usepackage[pagebackref,breaklinks,colorlinks,allcolors=cvprblue]{hyperref}
\usepackage{amsmath}  
\usepackage{amsthm}   
\usepackage[normalem]{ulem}
\usepackage{multirow}

\useunder{\uline}{\ul}{}
\usepackage{pifont}

\makeatletter
\newcommand\whline{\noalign{\ifnum0=`}\fi\hrule \@height 1.25pt \futurelet
	\reserved@a\@xhline}
\makeatother

\title{Synchronized Video-to-Audio Generation via\\ Mel Quantization-Continuum Decomposition}

\author{Juncheng Wang$^1$$^2$\thanks{Equal contribution. $^{\dag}$Project lead. $^{\ddag}$Correspondence.}
~ ~ Chao Xu$^2$$^*$
~ ~ Cheng Yu$^2$$^*$
~ ~ Lei Shang$^2$ $^{\dag}$
~ ~ Zhe Hu$^1$ \\
~ ~ Shujun Wang$^1$ $^{\ddag}$
~ ~ Liefeng Bo$^2$$^{\ddag}$\\
\normalsize $^1$ The Hong Kong Polytechnic University ~ $^2$Tongyi Lab, Alibaba Group }

\newtheorem{proposition}{Proposition}

\begin{document}
\maketitle

\begin{abstract}
Video-to-audio generation is essential for synthesizing realistic audio tracks that synchronize effectively with silent videos.
Following the perspective of extracting essential signals from videos that can precisely control the mature text-to-audio generative diffusion models, this paper presents how to balance the representation of mel-spectrograms in terms of completeness and complexity through a new approach called Mel Quantization-Continuum Decomposition (Mel-QCD).
We decompose the mel-spectrogram into three distinct types of signals, employing quantization or continuity to them, we can effectively predict them from video by a devised video-to-all (V2X) predictor.
Then, the predicted signals are recomposed and fed into a ControlNet, along with a textual inversion design, to control the audio generation process.
Our proposed Mel-QCD method demonstrates state-of-the-art performance across eight metrics, evaluating dimensions such as quality, synchronization, and semantic consistency. Our codes and demos will be released at \href{Website}{https://wjc2830.github.io/MelQCD/}.

\end{abstract}

\section{Introduction}
\label{sec:intro}

The relevance of multimedia content has positioned video-to-audio (V2A)~\cite{jeong2024read, xu2024video, xing2024seeing, zhang2024foleycrafter, iashin2021taming} generation as a critical area of research, focusing on synthesizing realistic audio tracks synchronized with silent video footage. This task is vital for enhancing user experiences in applications such as video editing~\cite{video_edit_1, video_edit_2}, post-production~\cite{post_prodct_1,post_prodct_2}, and content creation~\cite{hu2024animate, tian2024emo}, while also improving accessibility in media consumption. 
Moreover, as AI increasingly influences video production—often resulting in silent outputs~\cite{brooks2024video,SoRA,PiKA}—the development of effective V2A solutions has become essential. High-quality audio generation for AI-generated videos enhances storytelling and emotional impact, bridging the gap between visual and auditory information in various multimedia contexts.

\begin{figure}
    \centering
    \includegraphics[width=1.0\linewidth]{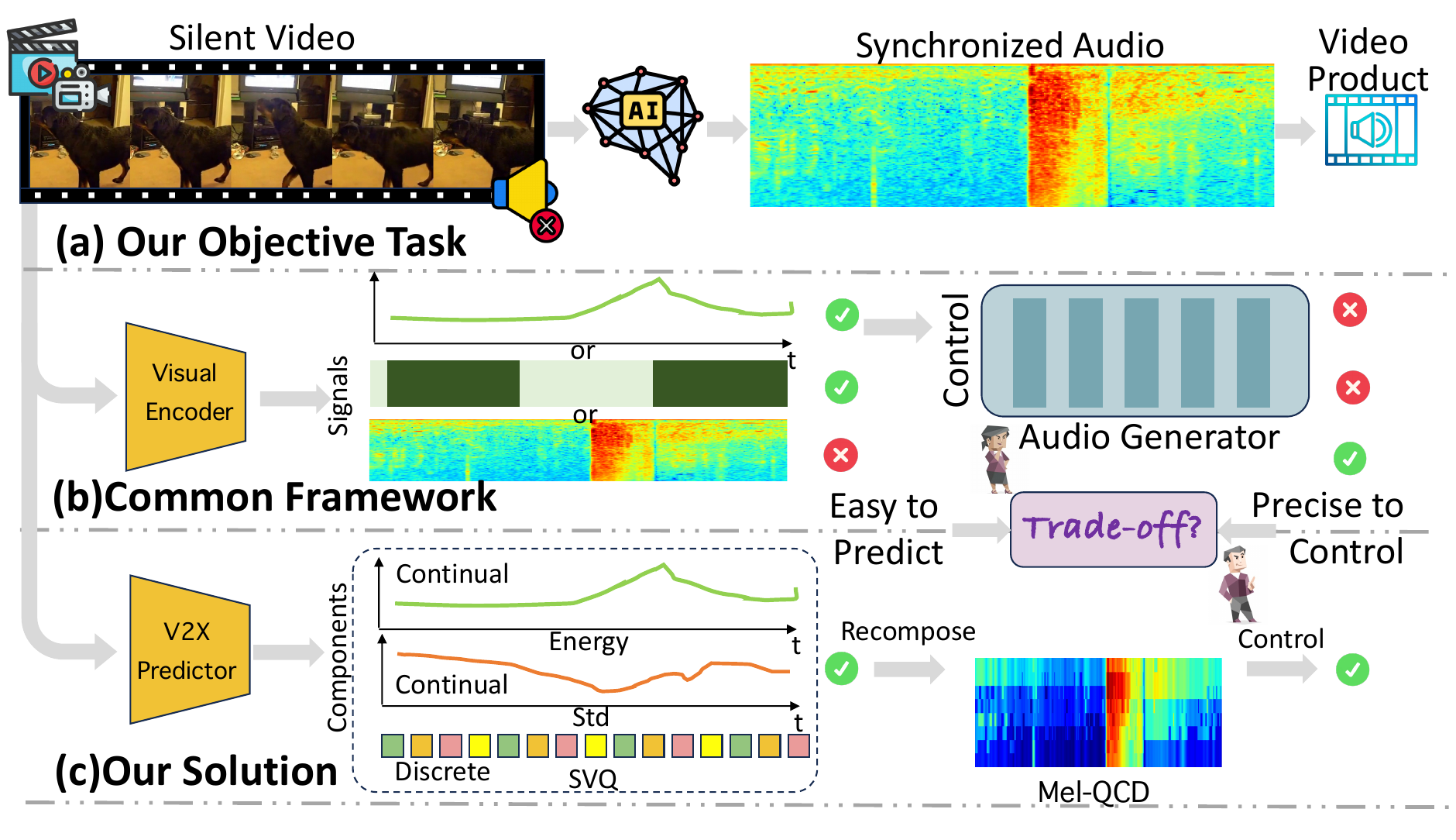}
    \caption{Task formulation of this paper (a). Previous mainstream approaches focus on extracting control signals from videos to govern audio generation (b). However, they struggle to balance between ease of prediction and precision of control. In response, our proposed Mel-QCD achieves a more effective trade-off (c).}
    \label{fig:teasor}
\end{figure}

Recent studies have shown promising results in V2A tasks by fine-tuning approaches from mature text-to-audio (T2A)~\cite{xue2024auffusion, liu2023audioldm, huang2023make, liu2024audioldm, huang2023make2} generation, particularly diffusion models. While these methods are driven by various motivations, they can generally be viewed as extracting signals from videos to control the synthesis of audio represented in mel-spectrogram format, as illustrated in Figure~\ref{fig:teasor}. For example, FoleyCrafter~\cite{zhang2024foleycrafter} extracts sound event onset signals from video, which are then used to control the temporal distribution within a text-to-audio generative diffusion model.

From this perspective, we hypothesize the primary challenge that the community faces is \textbf{how to better represent an audio signal that can be easily predicted from video while also enabling precise control over audio generation?}
Intuitively, an easily predictable signal promises effective generalization across different videos. However, to ensure that the generated audio is highly aligned with the conditioned video in both semantic and temporal representations, the signal must encompass adequate information. 
However, existing methods that fine-tune T2A diffusion models often fall short in one of these two dimensions, limitations that will be further discussed in Section~\ref{sec:related_work}.

In this paper, we present Mel Quantization-Continuum Decomposition (Mel-QCD) Control, as a way of trade-off between simplicity and precision.
Specifically, we approach the challenge by balancing the representation of the mel-spectrogram in terms of \emph{completeness} and \emph{complexity}.
We begin by decomposing the mel-spectrogram into three components: semantic, energy, and standard deviation vectors. 
To facilitate signal prediction from video, we propose quantizing the semantic vectors into discrete codes and constructing a semantic vector quantization (SVQ) codebook, where the index in the codebook represents the semantic component. This quantization effectively reduces \emph{complexity}, without losing too much information.
Moreover, for the energy and standard deviation vectors, as low-dimension data, we analyze their strong dependency on maintaining continuous distributions to ensure \emph{completeness}. Hence, we advocate for retaining their original representations.

By decomposing the mel-spectrogram into these components, we develop a video-to-all (V2X) predictor that generates signals from a given video. These individually predicted signals are subsequently recomposed to form a mel-like representation, termed Mel-QCD. Following FoleyCrafter~\cite{zhang2024foleycrafter} and ReWas~\cite{jeong2024read}, we employ a ControlNet~\cite{zhang2023adding} to manage the spatial distribution of the generated mel through a T2A diffusion model.
Furthermore, recognizing that the predicted Mel-QCD inevitably has some deviations in certain time slots, we incorporate a textual inversion technique to mitigate the semantic shifts caused by these inaccuracies. This enhancement ensures a more consistent alignment of the generated audio with the video, thereby preserving the fundamental semantic integrity.

A primary objective of this work is to explore a more efficient framework for generating high-quality audio conditioned by video. To evaluate our method, we compare its performance with existing mainstream pipelines on the widely recognized VGGSound benchmark across three perspectives: generation quality, synchronization, and semantic consistency, utilizing eight different metrics. Additionally, we conduct extensive analytical experiments to validate the proposed insights. 

\section{Related Work}
\label{sec:related_work}

\noindent\textbf{Text-to-Audio Generation}
The primary driver of success in audio generation tasks is the use of Text-to-Audio (TTA) models. Previous efforts adopt GANs~\cite{goodfellow2014generative, kreuk2022audiogen}, normalizing flows~\cite{papamakarios2021normalizing, kim2020glow}, and VAEs~\cite{van2017neural} in this task. Recently, diffusion models~\cite{ho2020denoising} have demonstrated significant generative potential in visual domains and have subsequently been extended to audio generation. DiffSound~\cite{yang2023diffsound} employs a discrete diffusion model to generate sounds from text prompts. Make-An-Audio~\cite{huang2023make} and AudioLDM~\cite{liu2023audioldm} excel in generation quality by leveraging advanced latent diffusion models (LDMs)~\cite{rombach2022high}. The following works further incorporates several enhancing designs. Make-An-Audio2~\cite{huang2023make2} introduces structured text encoding to improve semantic alignment. AudioLDM2~\cite{liu2024audioldm} unifies any audio signal into a universal representation, thus supporting various audio generation. Tango~\cite{ghosal2023text} utilizes Flan-T5~\cite{chung2024scaling} as the text encoder to precisely convey complex textual concepts. Auffusion~\cite{xue2024auffusion} reduces the demand for data and computational resources while maintaining excellent text-audio alignment. In this paper, our method is built upon Auffusion, inheriting its excellent text-conditioned generation capabilities while extending it to understand visual semantics.

\noindent\textbf{Video-to-Audio Generation}
Another popular research, Video-to-Audio (VTA), aims to generate audio that is semantically aligned and temporally synchronized with the video content. Based on how they connect visual and audio, we categorize the current work into two types. The first category involves training video-to-audio models from scratch. SpecVQGAN~\cite{iashin2021taming} and Im2wav~\cite{sheffer2023hear} design an autoregressive network to generate audio tokens based on the input visual tokens. DiffFoley~\cite{luo2024diff} employs a diffusion model while introducing contrastive learning to unify the video-audio features. VTA-LDM~\cite{xu2024video} is based on the LDMs and directly combine the visual embeddings through cross attention. However, they require a large-scale high-quality visual-audio aligned data for training, hence another category of methods has emerged, resorting to the foundational TTA models. Seeing-and-hearing~\cite{xing2024seeing} and VTA-Mapper~\cite{wang2024v2a} both project audio to the text embedding space, which is then processed by a pretrained TTA generator. However, the shared video-text space struggles to maintain the fine-grained temporal cues. Concurrent FoleyCrafter~\cite{zhang2024foleycrafter} and ReWaS~\cite{jeong2024read}, inspired by the structural control of ControlNet~\cite{zhang2023adding} in text-to-image tasks, utilize onset or energy inferred from visuals as mel-spectrogram hints to explicitly control audio generation. Apparently, these representations lose a lot of visual details, leading to unsatisfactory results. An intuitive solution is to increase signal information, but the difficulty of signal prediction also increases accordingly. Consequently, we delve into the properties of mel-spectrograms and propose several designs to make them easier to predict from visual features, 
achieving a balance between completeness and complexity.
\newcommand{\xuc}[1]{{\color{red}[xuc: #1]}}
\begin{figure*}[t]
    \centering
    \includegraphics[width=0.9\linewidth]{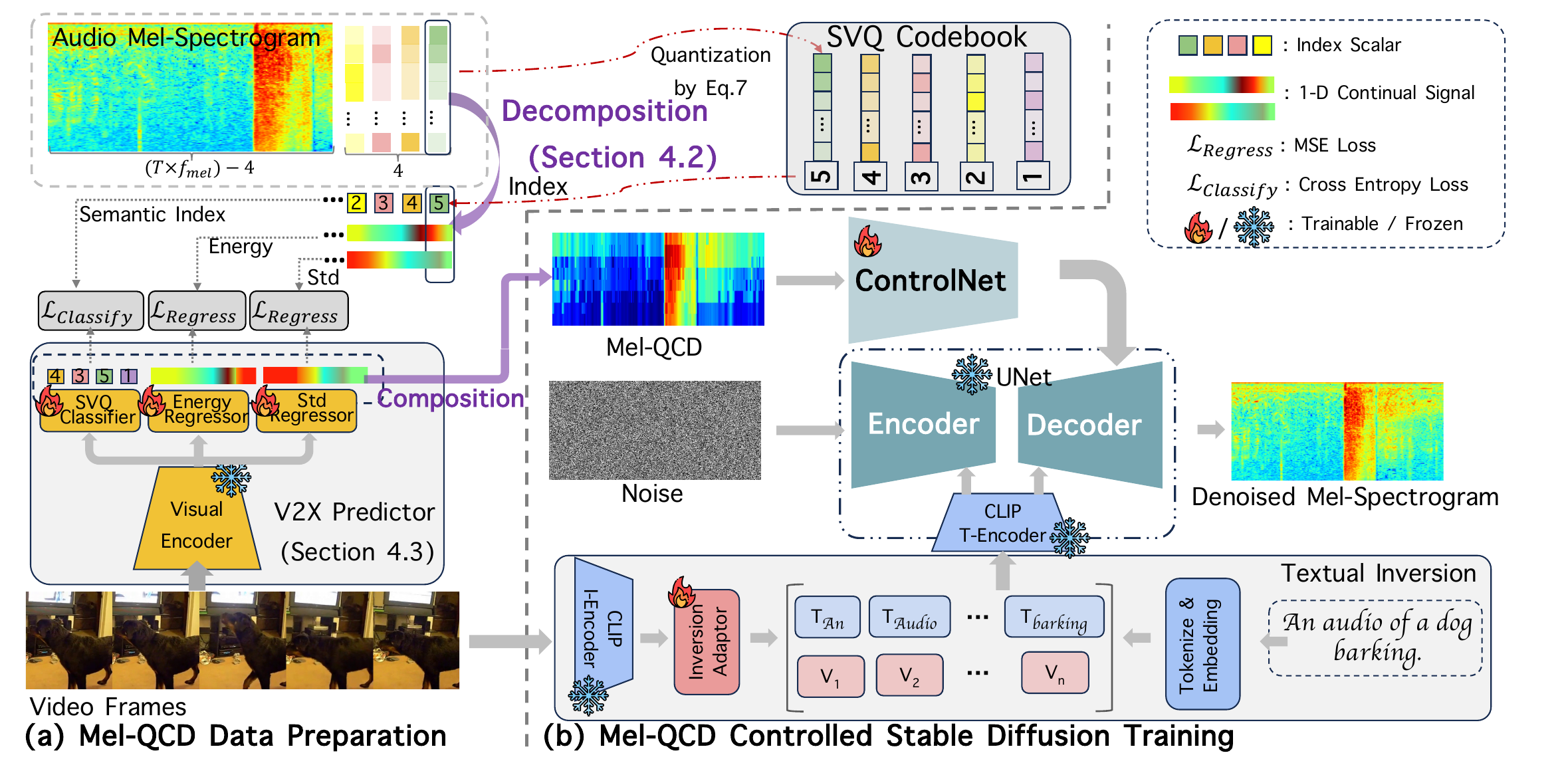}
    \caption{Pipeline for the proposed Mel-QCD controllable video-to-audio (V2A) generation. The process is divided into two parts: (a) Pre-training, which outlines how to derive Mel-QCD from videos; and (b) Training, which explains how to utilize Mel-QCD and textual inversion to train the video-controlled audio generation model.}
    \label{fig:pipeline}
\end{figure*}
\section{Task Formulation}
\label{sec:preliminary}
\noindent\textbf{Stable Diffusion Based Text-to-Audio} is one of the most popular T2A frameworks, where a $T$s length audio waveform $\mathbf{A}\in\mathbb{R}^{1\times(T\times f_{wav})}$ with a sample rate of $f_{wav}$ is transformed into a two dimensional Mel-Spectrogram $\mathbf{M}\in\mathbb{R}^{K\times(T\times f_{mel})}$ with a sample rate of $f_{mel}$, and a number of frequency bin of $K$, through a Fast Fourier Transform (FFT).
Previous T2A latent diffusion models (LDM)~\cite{xue2024auffusion, liu2023audioldm, liu2024audioldm} use paired audio $\mathbf{M}$ and textual embedding $\mathbf{C}_T$ data to train stable diffusion architecture models.
At inference time, the trained LDM $\mathcal{F}(\cdot;\theta)$ with corresponding parameters $\theta$, can transform the user prompted text feature $\mathbf{C}_T$ into a novel Mel-spectrogram as:
\begin{equation}
    \mathbf{M} = \mathcal{F}(\mathbf{C}_T;\theta).
\end{equation}
Later, with a synthesized Mel-spectrogram, a HiFi-GAN Vocoder~\cite{kong2020hifi} is used to transform it into audio waveform.

\noindent\noindent\textbf{ControlNet Based Video-to-Audio Generation}
ControlNet~\cite{zhang2023adding} complements the shortcoming of textual prompt that cannot control fine-grained spatial distribution of generated content. Formally, besides the textual prompt, ControlNet requires user to input an additional condition map $\mathbf{C}_S$, which describes the basic spatial distribution of the generated content. This explicit and fine-grained maps inspire video-to-audio generation methods~\cite{zhang2024foleycrafter, jeong2024read} to follow suit by converting videos into control hints, thereby achieving precise alignment. Specifically, given $T\times f_{v}$ frame video data $\mathbf{V}\in\mathbb{R}^{(T\times f_{v})\times 3\times H\times W}$, with a sample rate of $f_{v}$, frame resolution of $H\times W$, a signal predictor $\mathcal{G}$ transform them into the $\mathbf{C}_S=\mathcal{G}(\mathbf{V})\in\mathbb{R}^{K\times (T\times f_{mel})}$ of ControlNet with parameters $\Theta$. This process can be defined as:
\begin{equation}
    \mathbf{M} = \mathcal{F}(\mathbf{C}_S,\mathbf{C}_T;\Theta).
\end{equation}

In this study, we aim to address the challenge of generating well-aligned synthesized audio from video input. A crucial aspect of this endeavor is the effective prediction of a high-quality mel-spectrogram representation derived from the video. Consequently, we pose a central question that guides our research: \noindent\textbf{How can we achieve a better balance between the completeness and complexity of the controlling signal $\mathcal{G}(\mathbf{V})$ in video-to-audio (V2A) generation?} This inquiry underscores the significance of our work in enhancing the fidelity and synchronization of audio synthesis with corresponding video content.

\section{Our Approach}
\label{sec:approach}
\noindent\textbf{Overview:} To address the primary question posed earlier, we begin with a thorough analysis of the mel-spectrogram in Section~\ref{subsec:mel-decomp}. Building on the insights gained from this analysis, we introduce a novel representation known as Mel Quantization-Continuum Decomposition (Mel-QCD), which is detailed in Section~\ref{subsec:mel-qcd}. Utilizing this innovative representation, we develop V2X signal predictors aimed at extracting Mel-QCD from video inputs, as described in Section~\ref{subsec:v2x}. Finally, we discuss the integration of Mel-QCD within a textual enhanced stable diffusion pipeline in Section~\ref{subsec:qcd_sd}. The whole pipeline of our proposed method has been illustrated in Figure~\ref{fig:pipeline}.

\subsection{Mel-Signal Decomposition}
\label{subsec:mel-decomp}
To better understand the mel-input, we first decompose the original mel-spectrogram signal and analyze the significance of each component of the mel-map. This decomposition allows us to retain the most significant aspects of the signal, ensuring its completeness, while reducing complexity by discarding the relatively trivial components.

To achieve this, we begin by decomposing the mel-signal. Given a mel-map \(\mathbf{M} \in \mathbb{R}^{K \times (T \times f_{mel})}\), for any time slot \(t\), the audio signal \(\mathbf{M}_{.,t} \in \mathbb{R}^{K \times 1}\) can be expressed as:
\begin{align}
\mathbf{M}_{k,t} & = \mathbf{E}_{t} + \mathbf{S}_{k,t} \times \mathbf{D}_{t}, \quad \text{where} \label{eq:decompose} \\
\mathbf{E}_t & = \frac{1}{K} \sum_{k=1}^{K} \mathbf{M}_{k,t} \quad \in \mathbb{R}^{1}, \label{eq:energy} \\
\mathbf{D}_t & = \sqrt{\frac{1}{K} \sum_{k=1}^{K} (\mathbf{M}_{k,t} - \mathbf{E}_{t})^2} \quad \in \mathbb{R}^{1}, \label{eq:std} \\
\mathbf{S}_{.,t} & = \text{Norm}(\mathbf{M}_{.,t}) \quad \in \mathbb{R}^{K \times 1}. \label{eq:semantic}
\end{align}

After decomposing the mel-map into these three components, we will examine the properties of each one to assess their significance in representing the original mel-map.

\begin{figure}[t]
    \centering
    \includegraphics[width=1.0\linewidth]{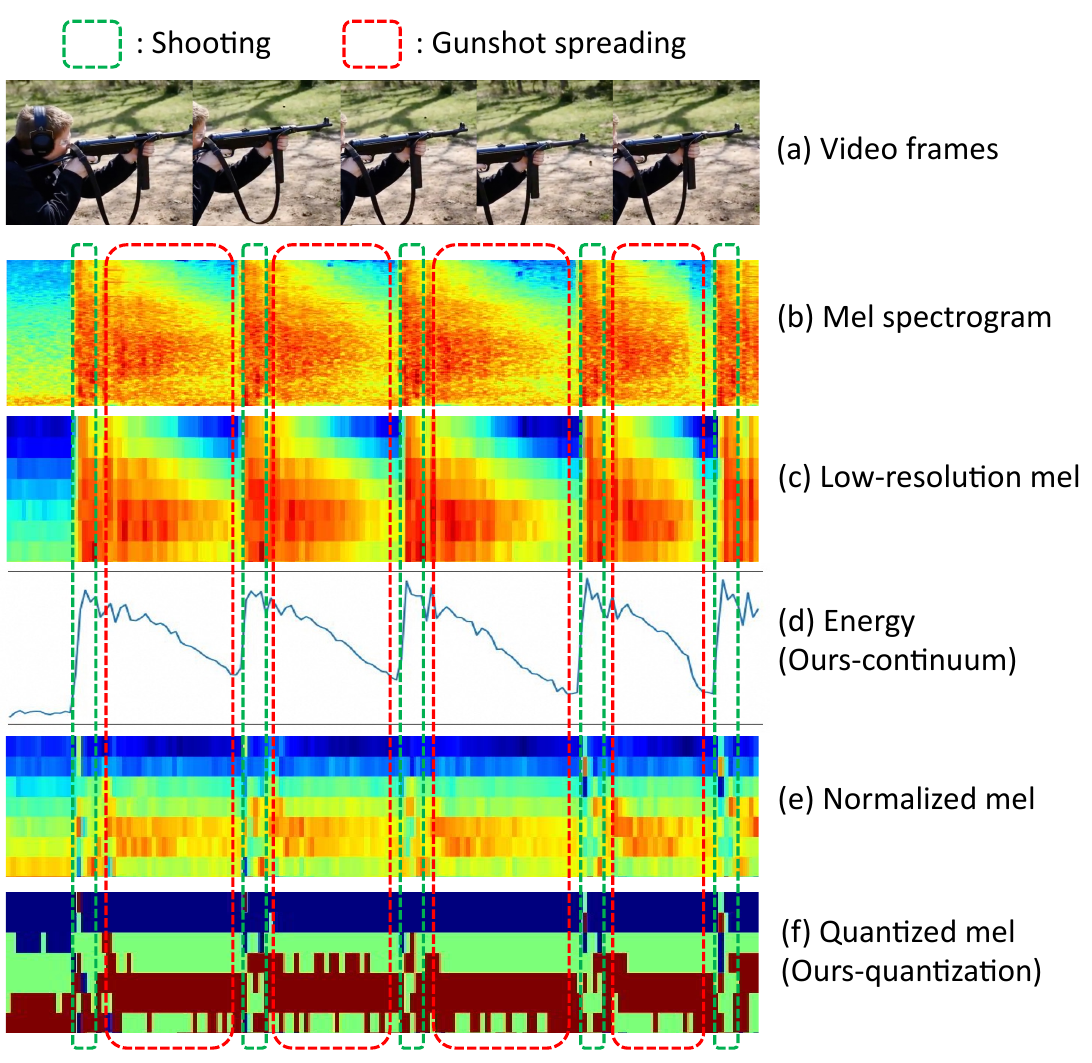}
    \caption{Properties of each component of the mel-map containing two sound events: shooting and gunshot spreading. The energy reflects the continuum of the mel-map while the normalized mel reflects the semantic clustering property.}
    \label{fig:decomposition}
\end{figure}
\begin{proposition}[Properties of Each Component]\label{prop:property}
    Given a mel-map represented audio \(\mathbf{M}\), the components \(\mathbf{S}_{.,t}\) tend to be distinguishable concerning sound events, while the other two components, \(\mathbf{E}_{t}\) and \(\mathbf{D}_{t}\), are continuously distributed across different sound events.
\end{proposition}
\noindent\textbf{Intuitive Explanation:} 
To begin with, \(\mathbf{S}_{.,t}\) represents the shape of the mel-distribution within a time slot. This shape, as described by~\cite{abbado1988perceptual}, determines the semantic content of the audio. Consequently, one can identify audio events based on their semantic characteristics. In contrast, \(\mathbf{E}_{t}\) reflects the energy of the sound, which tends to represent its overall loudness~\cite{jeong2024read}. As a result, sound events cannot be distinguished solely based on their loudness. Regarding \(\mathbf{D}_{t}\), it indicates the divergence in distribution across frequency dimensions, which includes semantic information~\cite{abbado1988perceptual}. However, its value distribution is heavily dependent on \(\mathbf{E}_{t}\), leading to a continuous distribution across sound events.

\noindent\textbf{Empirical Explanation:} 
Figure~\ref{fig:decomposition} demonstrates different components of the mel spectrogram for a video in which a man shoots continuously.
Generally, there exists two kinds of sound events, shooting and gunshot spreading.
As is shown in Figure~\ref{fig:decomposition}~(b)(c), both the original and low-resolution $\mathbf{M}_{.,t}$ in different time slots of gunshot spreading are not similar, thus cannot effectively cluster into a certain class for the sound event.
However, after our decomposition in Figure~\ref{fig:decomposition}(d)(e), $\mathbf{E}_{t}$ contains the main continuum of $\mathbf{M}_{.,t}$, and $\mathbf{S}_{.,t}$ has a better clustering property in the certain sound event.
As is shown in Figure~\ref{fig:decomposition}~(f), $\mathbf{S}_{.,t}$ can be quantized to decrease complexity with minimum loss of semantic information with the help of the clustering property.
As such, we can improve the trade-off of completeness and complexity by the above quantization-continuum decomposition, which will be elaborated in Section~\ref{subsec:mel-qcd}.

\subsection{Mel Quantization-Continuum Decomposition}\label{subsec:mel-qcd}

Although we use three components to represent the complete mel-map, the difficulty in inferring \(\mathbf{S}_{.,t}\) from videos should not be overlooked. To balance the signal completeness and predicting complexity, we present our proposed solution, which we term Mel Continuum-Quantization Decomposition (Mel-CQD). 
Specifically, considering the distinguishable nature of sound events in the semantic component \(\mathbf{S}_{.,t}\), the clustering distribution of these events allows us to quantize them while retaining the energy \(\mathbf{E}_t\) and standard deviation \(\mathbf{D}_t\) as continuous signals.

\noindent\textbf{Semantic Vector Quantization (SVQ).}
Given the clustering property of semantic vectors, the original continuous representation of \(\mathbf{S}_{.,t}\) assigns different indexes to vectors within the same cluster, resulting in redundancy. 

For instance, given \(N\) samples belonging to \(M\) categories, where \(N >> M\), the continuous indexing of all samples incurs a complexity of \(\mathcal{O}(N)\). However, by leveraging their clustering properties, we can discretize the indices for samples that share the same category, thereby reducing complexity to \(\mathcal{O}(M)\). 

Thus, we propose to discretize the values within \(\mathbf{S}_{.,t}\). One intuitive approach is to cluster the values and replace them with their respective clustering centroids. However, due to the high complexity of clustering algorithms, we suggest an alternative method to quantize the semantic vector \(\mathbf{S} \mapsto [\mathbf{S}]\) by rounding and clamping its values. This process can be mathematically described as follows:
\begin{equation}
    [\mathbf{S}]_{k,t} = \begin{cases}
      \text{round}(\mathbf{S}_{k,t}) & \text{if } \mathbf{S}_{k,t} \in [-\lambda,\lambda] \\
      -\lambda & \text{if } \mathbf{S}_{k,t} < -\lambda \\
      \lambda & \text{if } \mathbf{S}_{k,t} > \lambda,
    \end{cases}
\end{equation}
where \(\lambda\) is a pre-defined bounding positive integer. This quantization strategy ensures that, compared to the continuous semantic vector, the quantized version does not lose significant completeness, thanks to its inherent intra-sound event compactness. However, it is vital to note, at each time slot, the dimension of \([\mathbf{S}]_{.,t}\) remains \(K \times 1\) meaning that quantization alone cannot eliminate prediction difficulty.

To address this, we propose constructing a codebook for the semanticly quantized vectors. Through quantization, the possible values of semantic vector become finite within any given time slot. This allows us to treat SVQ prediction as a classification task, selecting from a finite set of possibilities.
Specifically, for a SVQ at one time slot, $[\mathbf{S}]_{.,t}$, we consider all possible combinations of $K$ scalar values within a range of $[-\lambda, \lambda]$. This generates a codebook with a length of $(2\lambda+1)^K$, representing all feasible combinations. Instead of predicting $K$ individual items, we perform a single classification task over this set of $(2\lambda+1)^K$ codes. Consequently, this approach reduces the prediction complexity by a factor of $K$.

\noindent\textbf{Energy \& Standard Deviation Continuum.}
As previously discussed, the energy signal must be maintained in a continuous form to ensure precise representation. Therefore, we do not apply any special processing to it, opting instead to regress it directly from videos. Additionally, to reconstruct a mel-map, as derived in Eq.~\ref{eq:decompose}, the standard deviation \(\mathbf{D}\) is also essential. Given that its complexity is similar to that of the energy signal, we likewise retain it in a continuous format and directly regress it.
\subsection{V2X Signal Predictors}
\label{subsec:v2x}
With thoroughly processed ground-truth signals derived from the original mel-maps, we are now ready to design the video-to-all signals (V2X) predictors \(\mathcal{G} = \mathcal{H} \times \mathcal{P}\), where \(\mathcal{H}\) represents the visual encoder~\cite{iashin2024synchformer} and \(\mathcal{P}\) denotes the signal-specific predictors.

For the design of \(\mathcal{H}\), we resample the video signal \(\mathbf{V} \in \mathbb{R}^{(T \times f_v) \times \ldots} \) to obtain \(\mathbf{V'} \in \mathbb{R}^{\frac{T \times f_{mel}}{4} \times \ldots}\) and utilize off-the-shelf visual encoders to derive \(\mathcal{H}_1(\mathbf{V'}) \in \mathbb{R}^{\frac{T \times f_{mel}}{4} \times d}\), where \(d\) denotes the frame embedding dimension.

The design of \(\mathcal{P}\) varies across different signals, which we will elaborate on below:

\noindent\textbf{SVQ Classifier.}
Recapping, the length of the SVQ codebook is \((2\lambda + 1)^K\), and this length influences the complexity of task. Empirically, the mel-map is derived from the waveform via FFT with a frequency bin of \(K = 256\), which results in an exponential increase in codebook length. 

To mitigate this issue, we downsample the mel-map along the frequency dimension, resulting in \(K' = 8\). For the choice of \(\lambda\), we set \(\lambda = 1\). It is worth noting that we have conducted ablation studies on this part in Section~\ref{sec:experiments} and demonstrated that even with such a high level of compression, we can still achieve good conditional generation. This further supports the conclusion of Proposition~\ref{prop:property}, which highlights the substantial redundancy of semantic vectors.

By this method, we reduce the length of the codebook to \(3^8 = 6561\). Although this is a significant reduction, it is still too large for classification. Therefore, we propose to factorize the \(3^8\) classification into two \(3^4 = 81\) classifications~\cite{cherniuk2023quantization}, further diminishing task complexity. To this end, we employ some transformer and MLP layers to classify.

\noindent\textbf{Energy \& Standard Deviation Regressor.}
To regress these two continuous signals, we directly apply multiple naive transformer and MLP layers to predict the continuous scalars for each time slot. For details regarding some experimental techniques, please refer to the Appendix.

\subsection{Mel-QCD Controlled Stable Diffusion}
\label{subsec:qcd_sd}
After obtaining three kinds of signals by $\hat{\mathbf{S}},\hat{\mathbf{E}},\hat{\mathbf{D}}=\mathcal{G}(\mathbf{V}')$, we can compose it into our final Mel-QCD $\mathbf{M}^{qcd}$ by:
\begin{equation}
    \mathbf{M}^{qcd}_{k,t} = \hat{\mathbf{E}}_{t} + \hat{\mathbf{S}}_{k,t}\times \hat{\mathbf{D}}_{t}.
\end{equation}

Hence, $\mathbf{M}^{qcd}$ already includes a rich set of easily obtainable local dynamic semantics, but we cannot ensure that every time slot is perfectly precise. These local inaccuracies can lead to shifts in the overall distribution, thus we further introduce a Textual Inversion~\cite{gal2022image, TI_xuc} module to refine the global semantics. Unlike the semantic adapter used in FoleyCrafter, textual inversion can enhance the visual understanding capabilities without affecting the frozen U-Net features, thus allowing it to be trained independently and merged with ControlNet with minimal impact on the main data flow. Specifically, we first predefine the prompt according to the sound event label, which is tokenized and mapped into the token embedding space by employing a CLIP embedding lookup module~\cite{radford2021learning}. Then, the Inversion Adapter averages the CLIP visual embeddings of videos along the time dimension and maps them into aligned pseudo-word token embeddings $\left\{\boldsymbol{V}_1,\dots,\boldsymbol{V}_{n}\right\}$, as shown in Figure~\ref{fig:pipeline}. Note that this module is forward-only and $n$ tokens are set to describe the sound event as comprehensively as possible from the visual inputs. Then, the two token embeddings are concatenated and sent to the CLIP text encoder, obtaining the semantically enhanced textual guidance $\boldsymbol{C}_T$. 

Finally, we obtain $\mathbf{M}^{qcd}$ as ControlNet hints $\boldsymbol{C}_{S}$ and textual embedding $\boldsymbol{C}_T$, both are incorporated with a Text-to-Audio base model to train $\Theta$.
\begin{equation}
    \mathcal{L}=\mathbb{E}_{\mathbf{z}_0,t,\mathbf{C}_S,\mathbf{C}_T,\epsilon\sim\mathcal{N}(0,1)}[||\epsilon-\epsilon_\theta(\mathbf{z}_t,t,\mathbf{C}_S,\mathbf{C}_T)||^2_2],
\end{equation}
where $\mathbf{z}_t$ is the noisy latent, $t$ is the diffusion step, $\epsilon_\theta$ is a denosing network.


\begin{table*}[t]
\centering
\caption{Comparison of our method with other state-of-the-art Foley generation models using the VGGSound test set. The highest performance among the various methods is highlighted in bold, while the second-best performance is underlined.}

\label{tab:main_vggsound}
\resizebox{.85\textwidth}{!}{
\begin{tabular}{c|ccc|ccc|cc}
\whline
             & \multicolumn{3}{c|}{Quality}                    & \multicolumn{3}{c|}{Synchronization}           & \multicolumn{2}{c}{Semantic}  \\ 
\multirow{-2}{*}{Method}             & FID$^\downarrow$            & MKL$^\downarrow$           & Class ACC$^\uparrow$      & W-Dis$^\downarrow$         & JS-Div$^\downarrow$        & Onset ACC$^\uparrow$      & IB-AA$^\uparrow$         & IB-AV$^\uparrow$         \\\whline
Auffusion~\cite{xue2024auffusion}          & 22.26          & 5.54          & 18.36          & 0.44          & {\ul 0.12}    & 22.30          & 0.35          & 0.22          \\
SpecVQGAN~\cite{iashin2021taming}          & 19.31         & 6.47          & 5.64          & 0.45          & 0.10    & 24.34          & 0.18          & 0.13          \\
Im2Wav~\cite{sheffer2023hear}             & 16.16          & 5.66          & 16.70          & 0.45          & 0.13          & 22.10          & 0.31          & 0.20          \\
DiffFoley~\cite{luo2024diff}          & 15.15          & 6.47          & 23.27          & 0.49          & 0.14          & 16.02          & 0.32          & 0.23          \\
VTA-LDM~\cite{xu2024video}            & {\ul 11.77}    & 4.72          & 27.72          & {\ul 0.37}    & \textbf{0.11} & \textbf{26.83} & 0.44          & 0.28          \\
Seeing-and-Hearing~\cite{xing2024seeing} & 20.32          & 6.08          & 10.56          & 0.68          & 0.17          & 20.49          & 0.43          & \textbf{0.38} \\
FoleyCrafter~\cite{zhang2024foleycrafter}       & 13.11          & {\ul 4.14}    & {\ul 31.54}    & 0.43          & 0.13          & 24.33          & {\ul 0.48}    & 0.29          \\
\hline
\rowcolor[HTML]{EFEFEF} 
Ours               & \textbf{11.73} & \textbf{2.96} & \textbf{45.91} & \textbf{0.33} & \textbf{0.11} & {\ul 25.42}    & \textbf{0.52} & {\ul 0.31}    \\ \whline
\end{tabular}
}
\end{table*}
\section{Experiments}
\label{sec:experiments}
\subsection{Settings}
\textbf{Dataset.} The main experiments are conducted in the VGGSound dataset~\cite{chen2020vggsound}.
We manually curate 56k videos with semantically aligned and temporally synchronized audio, using 55k for training and the remaining 1.1k for testing. 
For analysis and ablations, we utilize the AvSync15~\cite{zhang2024audio}, a high-quality subset of VGGSound, and adhere to its predefined splits for training and testing samples.  

\noindent\textbf{Baselines.} We compare our method with recent SOTA methods. Firstly, Auffusion~\cite{xue2024auffusion} is a text-to-audio method that serves as our base model. Secondly, SpecVQGAN~\cite{iashin2021taming}, Im2Wav~\cite{sheffer2023hear}, DiffFoley~\cite{luo2024diff}, and VTA-LDM~\cite{xu2024video} represent existing video-to-audio generation approaches, each of which is trained from the scratch. Thirdly, Seeing-and-Hearing~\cite{xing2024seeing} and FoleyCrafter~\cite{zhang2024foleycrafter} are another type of video-to-audio method that involve text-to-audio priors, with the latter following the similar technological approach as us. Note that we do not include ReWaS~\cite{jeong2024read} because its open-source code cannot reproduce the results as expected.

\noindent\textbf{Evaluation Metrics.} We focus on three key aspects to conduct objective evaluations. \textbf{Quality}: Drawing on previous studies~\cite{luo2024diff, wang2024v2a, xing2024seeing}, we utilize the Frechet Distance (FID)~\cite{heusel2017gans} and Mean KL Divergence (MKL)~\cite{iashin2021taming} as metrics. Additionally, Class ACC is employed to gauge the classification accuracy of the generated sounds. \textbf{Synchronization}: We report the Onset ACC as used in CondFoleyGen~\cite{du2023conditional} and further analyze the Onset Detection Function~\cite{bello2005tutorial, wang2024tiva} for both generated and ground truth audios, then computing the Wasserstein Distance (W-Dis) and Jensen-Shannon Divergence (JS-Div) between the distribution of the two to measure the temporal alignment. \textbf{Semantic}: To evaluate semantic relevance, we compare the ImageBind scores~\cite{girdhar2023imagebind} between the generated audio and both the corresponding video (IB-AV) and ground truth audio (IB-AA).

\subsection{Main Results}
\begin{figure*}[t]
    \centering
    \includegraphics[width=\linewidth]{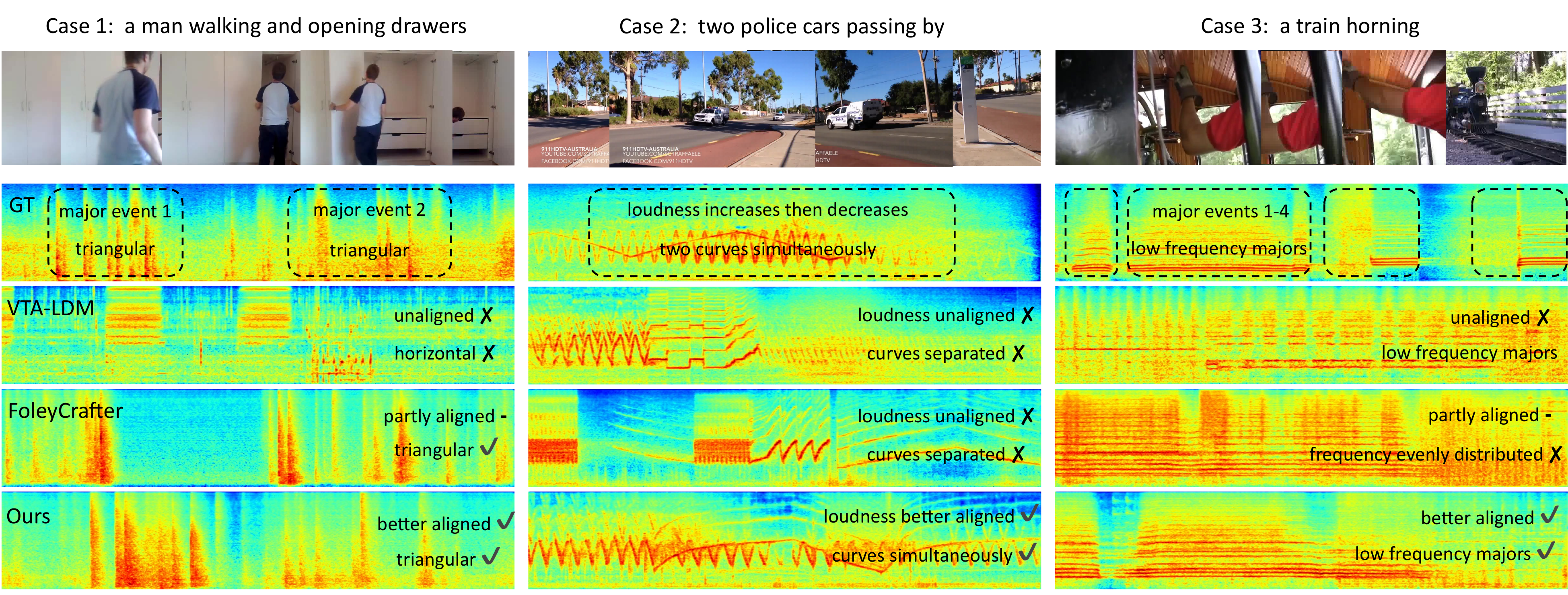}
    \caption{Case study on the VGGSound test set. The first row displays the video frame, followed by mel spectrograms from the ground truth (GT), VTA-LDM, FoleyCrafter, and our method. As shown, our result performs better than others over synchronization.}
    \label{fig:case_study_all}
\end{figure*}

As shown in Table~\ref{tab:main_vggsound}, we evaluate the results across three dimensions: generation quality, temporal synchronization, and semantic consistency. Notably, our method achieves state-of-the-art performance on the majority of metrics, with the exception of Onset ACC and IB-AV.
When comparing VTA-LDM to our approach, it exceeds our performance by approximately 1.5\% on the Onset Acc metric. Despite its marginally better performance, VTA-LDM exhibits equivalent performance to ours on the similar synchronization metric, JS-Div. We attribute this to the more training data that VTA-LDM utilizes, which likely allows it to rely on visual cues and prioritize the video reference, resulting in improved video-audio synchronization.
Regarding the IB-AV metric, it is important to note that Seeing-and-Hearing employs ImageBind as its visual encoder, thereby increasing the intrinsic relevance of its generated audio to the ImageBind video embeddings. Nonetheless, outside of Seeing-and-Hearing, our method still achieves state-of-the-art performance.
For qualitative evaluation, Figure~\ref{fig:case_study_all} presents a series of cases in the VGGSound test set.

\subsection{Comparison Among Different Control Signals}
\begin{table*}[]
\centering
\caption{Comparison among different controlling signals in terms of completeness and complexity. Results using ground-truth signals measure the completeness, while results using predicted signals measure their trade-off. The highest performance is highlighted in bold.}
\label{tab:com_controltype}
\resizebox{.95\textwidth}{!}{
\begin{tabular}{c|c|cccc|cccc}
\whline
\multirow{2}{*}{Control Type} & \multirow{2}{*}{Proposed by} & \multicolumn{4}{c|}{Ground Truth} & \multicolumn{4}{c}{Predicted} \\
& & FID$^\downarrow$   & MKL$^\downarrow$  & Class ACC$^\uparrow$ & Onset ACC$^\uparrow$ & FID$^\downarrow$ & MKL$^\downarrow$ & Class ACC$^\uparrow$ & Onset ACC$^\uparrow$ \\ \whline
\rowcolor[HTML]{EFEFEF} 
Ours                 & This paper   & {\ul47.57} & 1.67 & {\ul66.67}     & {\ul71.47} & \textbf{61.00} & \textbf{1.66} & \textbf{64.67} & 31.68 \\ \hline
Onset                & FoleyCrafter~\cite{zhang2024foleycrafter} & 65.38      &2.19     & 56.67          &63.31   &68.72 &2.34 &56.67 &27.62        \\ 
Energy               & ReWas~\cite{jeong2024read}        & 57.21 & 1.93 & 62.67     & 69.13  &71.18 &2.11  & 56.67 & 29.20  \\ 
Low-Resolution Mel   & TiVA~\cite{wang2024tiva}         & \textbf{47.50} & {\ul1.64} & 65.34     & 69.51  & {\ul65.10} & 2.64 & 52.00 & {\ul35.71}  \\ 
CLIP Video Embedding & EgoSonics~\cite{rai2024egosonics}    & 90.08 & 1.94 & 60.67     & 39.38  & 90.08 & {\ul1.94} & {\ul60.67} & \textbf{39.38}  \\ 
Codec-1              & EnCodec~\cite{defossez2022high}        & 60.00 & 1.81 & 62.00     & 70.09  & 102.80 & 3.31 & 40.67 & 30.92   \\ 
Codec-2              & EnCodec~\cite{defossez2022high}        & 48.42 & \textbf{1.39} & \textbf{69.34}     & \textbf{71.73}  & 101.13 & 3.09 & 40.67 & 29.67   \\ \whline
\end{tabular}

}

\end{table*}

We compare our proposed Mel-QCD with various audio-visual signals from recent research, including onset binary masks~\cite{zhang2024foleycrafter}, energy signals~\cite{jeong2024read}, low-resolution mel~\cite{wang2024tiva}, and CLIP video embeddings~\cite{rai2024egosonics}. Additionally, we evaluate our quantization against audio quantization, specifically the first (Codec-1) and first two (Codec-2) RVQ codebooks from \cite{defossez2022high}. And we maintain consistent setups, signal predictor and ControlNet, and without include textual inversion.

For each control signal type, we utilize both ground-truth and predicted measures to assess completeness and complexity. Ground-truth results gauge completeness, while predicted results quantify trade-offs. As shown in Table~\ref{tab:com_controltype}, Codec-2 exhibits the best overall performance in terms of completeness, effectively preserving original audio information. However, the EnCodec codebooks (Codec-1 and Codec-2) have high complexity due to their codebook sizes of $1024^3$ and $1024^6$.
Our proposed Mel-QCD performs comparably to low-resolution mel and significantly outperforms onset and energy, highlighting its capacity to retain semantic cues. CLIP video embeddings suffer the worst completeness and complexity since they derive from video rather than audio.
In terms of trade-offs, Codec-1 and Codec-2 all show relatively low performance, indicating that an excessive focus on either completeness or complexity undermines balance. In contrast, Mel-QCD achieves the best FID, MKL, and Class ACC, and ranks second in Onset ACC. Compared to onset and energy signals, Mel-QCD retains more semantics with minimal complexity increase. When compared to low-resolution mel, it retains nearly all semantic information while effectively reducing complexity. Consequently, Mel-QCD demonstrates superior performance in trade-off completeness and complexity.

\subsection{Ablation Study}
\definecolor{lightgray}{rgb}{0.9, 0.9, 0.9}
\definecolor{lgray}{rgb}{0.66, 0.66, 0.66}
\newcommand{\cmark}{\ding{51}\xspace}%
\newcommand{\cmarkg}{\textcolor{lgray}{\ding{51}}\xspace}%
\newcommand{\xmark}{\ding{55}\xspace}%
\newcommand{\xmarkg}{\textcolor{lgray}{\ding{55}}\xspace}%

\begin{table}[]
\centering
\caption{Ablation study on the method component.}

\label{tab:ablate_module}
\renewcommand{\arraystretch}{1.25}{
\resizebox{.47\textwidth}{!}{
\begin{tabular}{c|c|c|c|c|c|c}
\whline
MQ          & TI  & FID$^\downarrow$ & MKL$^\downarrow$  & Class ACC$^\uparrow$ & Onset ACC$^\uparrow$ & IB-AV$^\uparrow$\\ \whline
\xmarkg                   & \xmarkg & 126.60 & 4.66     & 43.33 & 23.17  & 27.36   \\ \hline
\cmark               & \xmarkg      & 61.00     & 1.66         & 64.67  & 31.68   & 31.32     \\ \hline
\xmarkg              & \cmark & 75.65 & 2.02     & 58.33  & 25.08 & 30.27 \\ \hline \rowcolor[HTML]{EFEFEF} 
\cmark          & \cmark & 57.36 & 1.53     & 69.33  & 34.46 & 33.24 \\ \whline
\end{tabular}

}}
\end{table}
\begin{figure}[t]
    \centering
    \includegraphics[width=1.0\linewidth]{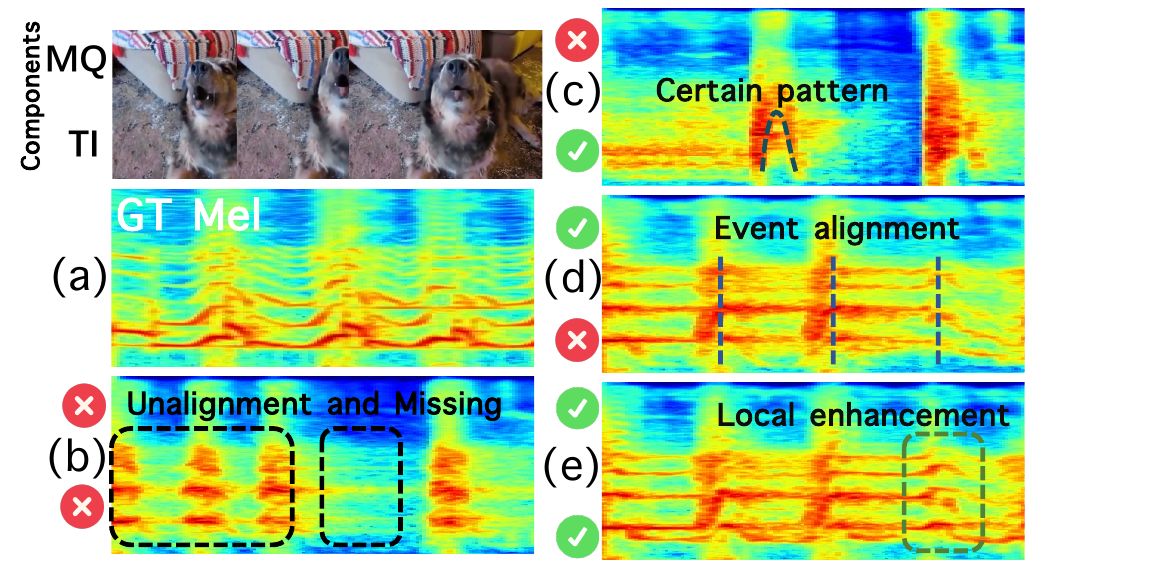}
    \caption{Visual comparison with different variants of our method.}
    \vspace{-0.5cm}
    \label{fig:abla_compo}
\end{figure}

\noindent\textbf{Ablating Method Component.}
To validate the effectiveness of the proposed Mel-QCD (MQ) and Textual Inversion (TI), we perform the following results. As illustrated in Figure~\ref{fig:abla_compo}, the ground truth (GT) mel-spectrogram (a) shows three dog barking events, each characterized by distinct frequency spikes. In the absence of MQ and TI, our base model (b) produces irregularly timed dog barks with visual-unrelated patterns.
When TI is applied alone (c), the model captures certain frequency patterns through global semantic control, though it lacks temporal alignment. Conversely, the MQ (d) provides explicit dynamic semantic cues, allowing the model to align sound events and fit local frequency spikes. Finally, incorporating TI alongside MQ (e) enhances local semantic details, resulting in more accurate overall distributions.
Quantitative results in Table~\ref{tab:ablate_module} support these claims that both modules improve performance.

\noindent\textbf{Ablating SVQ from Generation.}
\begin{table}[]
\centering
\caption{Ablation study on the choice of SVQ parameters in terms of the effectiveness of signal prediction and overall performance.}

\label{tab:ablate_cblength}
\renewcommand{\arraystretch}{1.05}{
\resizebox{.5\textwidth}{!}{
\begin{tabular}{cc|c|c|c|c|c}
\whline
\multicolumn{2}{c|}{Parameters} & \multirow{2}{*}{Codebook Length} & \multirow{2}{*}{ACC$^\uparrow$} & \multirow{2}{*}{Sim$^\uparrow$} & \multirow{2}{*}{FID$^\downarrow$} & \multirow{2}{*}{MKL$^\downarrow$}  \\ \cline{1-2}
\multicolumn{1}{c|}{$\lambda$}       & $K$      &                                  &                      &                      &                            &                          \\ \whline
\rowcolor[HTML]{EFEFEF} 
\multicolumn{1}{c|}{1}            & 8       & 6,561                            & 32.95 & 74.21 & 54.46 & 1.92 \\ \hline
\multicolumn{1}{c|}{1}            & 16      & 6,561$^2$         & 34.73 & 69.83 & 53.09 & 1.94                    \\ \hline
\multicolumn{1}{c|}{2}            & 16      & 390,625$^2$       & 20.90 & 71.39 & 55.10 & 2.27                    \\ \hline
\multicolumn{1}{c|}{2}            & 32      & 390,625$^4$       & 22.91 & 66.91 & 86.81 & 3.51                    \\ \whline
\end{tabular}
}
}
\end{table}
We assess the trade-off between completeness and complexity in our Semantic Vector Quantization (SVQ) by varying $\lambda$ and $K$ to adjust codebook length. We analyze classification accuracy (ACC), cosine similarity (Sim) of predictions, and quality metrics of the generated audio, using GT energy and standard deviation to isolate the impact of other control signals.

As shown in Table~\ref{tab:ablate_cblength}, decreases in ACC and Sim suggest that longer codebook lengths complicate signal prediction. Additionally, FID and MKL scores remain similar for the first two rows of the table but decline with increased codebook lengths, highlighting that the rising prediction difficulty significantly affects overall generation performance.

\noindent\textbf{Ablating Temporal Downsample Strategies.}
\begin{table}[]
\centering
\caption{Ablation study on the choice of down sampling strategy over the temporal dimension.}

\label{tab:ablate_temp_ds}
\renewcommand{\arraystretch}{1.05}{
\resizebox{.5\textwidth}{!}{

\begin{tabular}{c|c|c|c|c|c}
\whline
Temporal Down   & Number    & \multirow{2}{*}{FID$^\downarrow$} & \multirow{2}{*}{MKL$^\downarrow$} & \multirow{2}{*}{Class ACC$^\uparrow$} & \multirow{2}{*}{Onset ACC$^\uparrow$} \\
Sample Strategy & of Signal &                      &                      &                            &                            \\ \whline
\rowcolor[HTML]{EFEFEF} 
DownSample      & 256          & 47.87                & 1.65                 & 67.33                      & 69.10                      \\ \hline
Original        & 1024          & 47.54                & 1.67                 & 66.67                      & 71.47                      \\ \hline
Temoral Smooth  & 256          & 52.43                & 1.64                 & 62.67                      & 71.33                      \\ \hline
Temporal Mean   & 256           & 53.48                & 1.54                 & 61.33                      & 63.70                      \\ \whline
\end{tabular}
}}
\end{table}
As discussed in Section~\ref{sec:approach}, the number of video frames is less than the temporal length of the original mel spectrogram. Consequently, we cannot predict the number of signals that matches the temporal length of the mel. For instance, at a frame rate of 25 frames per second (FPS), a typical video would generate 250 frames, while the mel spectrogram has a temporal length of 1024. This necessitates downsampling to achieve a temporal dimension reduction by a factor of $\frac{1}{4}$.

To evaluate the choice of downsampling strategies, we compare three variants: downsampling with repetition, smoothing using the Savitzky-Golay filter~\cite{scipy}, and downsampling by taking the mean. As shown in Table~\ref{tab:ablate_temp_ds}, the naive approach of downsampling with repetition yields results comparable to those using the original resolution.

\noindent\textbf{Ablating Frequency Downsample Strategies.}
\begin{table}[]
\centering
\caption{Ablation study on the choice of compression strategy over the frequency dimension.}

\label{tab:ablate_freq_ds}
\renewcommand{\arraystretch}{1.05}{
\resizebox{.45\textwidth}{!}{
\begin{tabular}{c|c|c|c|c}
\whline
Frequency Down  &                       &                       &                             &                             \\
Sample Strategy & \multirow{-2}{*}{FID$^\downarrow$} & \multirow{-2}{*}{MKL$^\downarrow$} & \multirow{-2}{*}{Class ACC$^\uparrow$} & \multirow{-2}{*}{Onset ACC$^\uparrow$} \\ \whline
\rowcolor[HTML]{EFEFEF} 
DS-Repeat       & 47.57                 & 1.67                  & 66.67                       & 71.47                       \\ \hline
DS-Sparse       & 50.04                 & 1.68                  & 64.00                       & 70.42                       \\ \hline
Mel-8 Repeat    & 51.50                 & 1.72                  & 59.33                       & 69.10                       \\ \hline
Mel-8 Sparse    & 58.01                 & 2.10                  & 56.67                       & 70.25                       \\ \whline
\end{tabular}
}}
\end{table}
For the compression of the frequency dimension, we employ two main strategies: downsampling and performing a Fast Fourier Transform (FFT) with a reduced number of frequency bins. To achieve recovery that meets the resolution requirements of ControlNet, we consider two approaches: repetition and sparsity. As demonstrated in Table~\ref{tab:ablate_freq_ds}, the straightforward method of downsampling with repetition remains the most effective option.

\noindent\textbf{Ablating the Number of Pseudo-Word Tokens.}
\begin{table}[]
\centering
\caption{Ablation study on the number of pseudo-word tokens.}

\label{tab:ablate_titoken}
\renewcommand{\arraystretch}{1}{
\resizebox{.34\textwidth}{!}{
\begin{tabular}{c|c|c|c|c}
\whline
$n$          & FID$^\downarrow$   & MKL$^\downarrow$  & Class ACC$^\uparrow$ & IB-AV$^\uparrow$ \\ \whline
1                   & 110.21 & 4.50 & 35.33     & 27.85     \\ \hline
4               &  92.95     & 3.03     & 47.67         & 28.62          \\ \hline
16                     & 83.32 & 2.61 & 54.33     & 29.17    \\ \hline\rowcolor[HTML]{EFEFEF} 
32          & 75.65 & 2.02 & 58.33     & 30.27     \\ \hline
64     & 79.18 & 1.91 & 55.67     & 30.30    \\ \whline
\end{tabular}

}}
\end{table}
To find the optimal number of pseudo-word tokens, we assess audio quality and semantic alignment for different values $n$. As shown in Table~\ref{tab:ablate_titoken}. It is clear that as $n$ increases, the overall semantics becomes more detailed, reflecting on the improvement of various metrics. However, when increasing $n$ from 32 to 64, there is no significant boosting in overall performance. Considering increase the number of pseudo-word tokens raises memory usage, the optimal balance between computational load and performance is at $n=32$.
\section{Conclusion}
\label{sec:conclusion}
In this paper, we introduced a novel method for generating audio that aligns closely with conditional videos using Mel Quantization-Continuum Decomposition (Mel-QCD). We emphasized that representing the mel-spectrogram with predictable and comprehensive information enables effective control over text-to-audio diffusion models, facilitating seamless audio synthesis that corresponds to the video. Mel-QCD aims to optimize the trade-off between completeness and complexity in mel representation, further enhanced by our video-to-all (V2X) predictor for precise audio generation control.
Extensive experiments and ablation studies demonstrate that our approach produces high-quality audio with superior temporal synchronization and semantic consistency when conditioned on video input.

\noindent\textbf{Limitations} We recognize that our experimental evaluation is constrained by computational limitations and the difficulty of acquiring large datasets. Future work will aim to advance industrial-level model training to improve the applicability and scalability of our approach.
{
    \small
    \bibliographystyle{ieeenat_fullname}
    \bibliography{ref}

\begin{thebibliography}{50}
\providecommand{\natexlab}[1]{#1}
\providecommand{\url}[1]{\texttt{#1}}
\expandafter\ifx\csname urlstyle\endcsname\relax
  \providecommand{\doi}[1]{doi: #1}\else
  \providecommand{\doi}{doi: \begingroup \urlstyle{rm}\Url}\fi

\bibitem[Abbado(1988)]{abbado1988perceptual}
Adriano Abbado.
\newblock Perceptual correspondences of abstract animation and synthetic sound.
\newblock \emph{Leonardo. Supplemental Issue}, pages 3--5, 1988.

\bibitem[Bello et~al.(2005)Bello, Daudet, Abdallah, Duxbury, Davies, and Sandler]{bello2005tutorial}
Juan~Pablo Bello, Laurent Daudet, Samer Abdallah, Chris Duxbury, Mike Davies, and Mark~B Sandler.
\newblock A tutorial on onset detection in music signals.
\newblock \emph{IEEE Transactions on speech and audio processing}, 13\penalty0 (5):\penalty0 1035--1047, 2005.

\bibitem[Brooks et~al.(2024)Brooks, Peebles, Holmes, DePue, Guo, Jing, Schnurr, Taylor, Luhman, Luhman, et~al.]{brooks2024video}
Tim Brooks, Bill Peebles, Connor Holmes, Will DePue, Yufei Guo, Li Jing, David Schnurr, Joe Taylor, Troy Luhman, Eric Luhman, et~al.
\newblock Video generation models as world simulators. 2024.
\newblock \emph{URL https://openai. com/research/video-generation-models-as-world-simulators}, 3, 2024.

\bibitem[Ceylan et~al.(2023)Ceylan, Huang, and Mitra]{video_edit_1}
Duygu Ceylan, Chun-Hao~P Huang, and Niloy~J Mitra.
\newblock Pix2video: Video editing using image diffusion.
\newblock In \emph{Proceedings of the IEEE/CVF International Conference on Computer Vision}, pages 23206--23217, 2023.

\bibitem[Chen et~al.(2020)Chen, Xie, Vedaldi, and Zisserman]{chen2020vggsound}
Honglie Chen, Weidi Xie, Andrea Vedaldi, and Andrew Zisserman.
\newblock Vggsound: A large-scale audio-visual dataset.
\newblock In \emph{ICASSP 2020-2020 IEEE International Conference on Acoustics, Speech and Signal Processing (ICASSP)}, pages 721--725. IEEE, 2020.

\bibitem[Cherniuk et~al.(2023)Cherniuk, Abukhovich, Phan, Oseledets, Cichocki, and Gusak]{cherniuk2023quantization}
Daria Cherniuk, Stanislav Abukhovich, Anh-Huy Phan, Ivan Oseledets, Andrzej Cichocki, and Julia Gusak.
\newblock Quantization aware factorization for deep neural network compression.
\newblock \emph{arXiv preprint arXiv:2308.04595}, 2023.

\bibitem[Chung et~al.(2024)Chung, Hou, Longpre, Zoph, Tay, Fedus, Li, Wang, Dehghani, Brahma, et~al.]{chung2024scaling}
Hyung~Won Chung, Le Hou, Shayne Longpre, Barret Zoph, Yi Tay, William Fedus, Yunxuan Li, Xuezhi Wang, Mostafa Dehghani, Siddhartha Brahma, et~al.
\newblock Scaling instruction-finetuned language models.
\newblock \emph{Journal of Machine Learning Research}, 25\penalty0 (70):\penalty0 1--53, 2024.

\bibitem[D{\'e}fossez et~al.(2022)D{\'e}fossez, Copet, Synnaeve, and Adi]{defossez2022high}
Alexandre D{\'e}fossez, Jade Copet, Gabriel Synnaeve, and Yossi Adi.
\newblock High fidelity neural audio compression.
\newblock \emph{arXiv preprint arXiv:2210.13438}, 2022.

\bibitem[Du et~al.(2023)Du, Chen, Salamon, Russell, and Owens]{du2023conditional}
Yuexi Du, Ziyang Chen, Justin Salamon, Bryan Russell, and Andrew Owens.
\newblock Conditional generation of audio from video via foley analogies.
\newblock In \emph{Proceedings of the IEEE/CVF Conference on Computer Vision and Pattern Recognition}, pages 2426--2436, 2023.

\bibitem[Gal et~al.(2022)Gal, Alaluf, Atzmon, Patashnik, Bermano, Chechik, and Cohen-Or]{gal2022image}
Rinon Gal, Yuval Alaluf, Yuval Atzmon, Or Patashnik, Amit~H Bermano, Gal Chechik, and Daniel Cohen-Or.
\newblock An image is worth one word: Personalizing text-to-image generation using textual inversion.
\newblock \emph{arXiv preprint arXiv:2208.01618}, 2022.

\bibitem[Ghosal et~al.(2023)Ghosal, Majumder, Mehrish, and Poria]{ghosal2023text}
Deepanway Ghosal, Navonil Majumder, Ambuj Mehrish, and Soujanya Poria.
\newblock Text-to-audio generation using instruction-tuned llm and latent diffusion model.
\newblock \emph{arXiv preprint arXiv:2304.13731}, 2023.

\bibitem[Girdhar et~al.(2023)Girdhar, El-Nouby, Liu, Singh, Alwala, Joulin, and Misra]{girdhar2023imagebind}
Rohit Girdhar, Alaaeldin El-Nouby, Zhuang Liu, Mannat Singh, Kalyan~Vasudev Alwala, Armand Joulin, and Ishan Misra.
\newblock Imagebind: One embedding space to bind them all.
\newblock In \emph{Proceedings of the IEEE/CVF Conference on Computer Vision and Pattern Recognition}, pages 15180--15190, 2023.

\bibitem[Goodfellow et~al.(2014)Goodfellow, Pouget-Abadie, Mirza, Xu, Warde-Farley, Ozair, Courville, and Bengio]{goodfellow2014generative}
Ian Goodfellow, Jean Pouget-Abadie, Mehdi Mirza, Bing Xu, David Warde-Farley, Sherjil Ozair, Aaron Courville, and Yoshua Bengio.
\newblock Generative adversarial nets.
\newblock \emph{Advances in neural information processing systems}, 27, 2014.

\bibitem[Heusel et~al.(2017)Heusel, Ramsauer, Unterthiner, Nessler, and Hochreiter]{heusel2017gans}
Martin Heusel, Hubert Ramsauer, Thomas Unterthiner, Bernhard Nessler, and Sepp Hochreiter.
\newblock Gans trained by a two time-scale update rule converge to a local nash equilibrium.
\newblock \emph{Advances in neural information processing systems}, 30, 2017.

\bibitem[Ho et~al.(2020)Ho, Jain, and Abbeel]{ho2020denoising}
Jonathan Ho, Ajay Jain, and Pieter Abbeel.
\newblock Denoising diffusion probabilistic models.
\newblock \emph{Advances in neural information processing systems}, 33:\penalty0 6840--6851, 2020.

\bibitem[Hu(2024)]{hu2024animate}
Li Hu.
\newblock Animate anyone: Consistent and controllable image-to-video synthesis for character animation.
\newblock In \emph{Proceedings of the IEEE/CVF Conference on Computer Vision and Pattern Recognition}, pages 8153--8163, 2024.

\bibitem[Huang et~al.(2023{\natexlab{a}})Huang, Ren, Huang, Yang, Ye, Zhang, Liu, Yin, Ma, and Zhao]{huang2023make2}
Jiawei Huang, Yi Ren, Rongjie Huang, Dongchao Yang, Zhenhui Ye, Chen Zhang, Jinglin Liu, Xiang Yin, Zejun Ma, and Zhou Zhao.
\newblock Make-an-audio 2: Temporal-enhanced text-to-audio generation.
\newblock \emph{arXiv preprint arXiv:2305.18474}, 2023{\natexlab{a}}.

\bibitem[Huang et~al.(2023{\natexlab{b}})Huang, Huang, Yang, Ren, Liu, Li, Ye, Liu, Yin, and Zhao]{huang2023make}
Rongjie Huang, Jiawei Huang, Dongchao Yang, Yi Ren, Luping Liu, Mingze Li, Zhenhui Ye, Jinglin Liu, Xiang Yin, and Zhou Zhao.
\newblock Make-an-audio: Text-to-audio generation with prompt-enhanced diffusion models.
\newblock In \emph{International Conference on Machine Learning}, pages 13916--13932. PMLR, 2023{\natexlab{b}}.

\bibitem[Iashin and Rahtu(2021)]{iashin2021taming}
Vladimir Iashin and Esa Rahtu.
\newblock Taming visually guided sound generation.
\newblock \emph{arXiv preprint arXiv:2110.08791}, 2021.

\bibitem[Iashin et~al.(2024)Iashin, Xie, Rahtu, and Zisserman]{iashin2024synchformer}
Vladimir Iashin, Weidi Xie, Esa Rahtu, and Andrew Zisserman.
\newblock Synchformer: Efficient synchronization from sparse cues.
\newblock In \emph{ICASSP 2024-2024 IEEE International Conference on Acoustics, Speech and Signal Processing (ICASSP)}, pages 5325--5329. IEEE, 2024.

\bibitem[Jeong et~al.(2024)Jeong, Kim, Chun, and Lee]{jeong2024read}
Yujin Jeong, Yunji Kim, Sanghyuk Chun, and Jiyoung Lee.
\newblock Read, watch and scream! sound generation from text and video.
\newblock \emph{arXiv preprint arXiv:2407.05551}, 2024.

\bibitem[Kim et~al.(2020)Kim, Kim, Kong, and Yoon]{kim2020glow}
Jaehyeon Kim, Sungwon Kim, Jungil Kong, and Sungroh Yoon.
\newblock Glow-tts: A generative flow for text-to-speech via monotonic alignment search.
\newblock \emph{Advances in Neural Information Processing Systems}, 33:\penalty0 8067--8077, 2020.

\bibitem[Kong et~al.(2020)Kong, Kim, and Bae]{kong2020hifi}
Jungil Kong, Jaehyeon Kim, and Jaekyoung Bae.
\newblock Hifi-gan: Generative adversarial networks for efficient and high fidelity speech synthesis.
\newblock \emph{Advances in neural information processing systems}, 33:\penalty0 17022--17033, 2020.

\bibitem[Kreuk et~al.(2022)Kreuk, Synnaeve, Polyak, Singer, D{\'e}fossez, Copet, Parikh, Taigman, and Adi]{kreuk2022audiogen}
Felix Kreuk, Gabriel Synnaeve, Adam Polyak, Uriel Singer, Alexandre D{\'e}fossez, Jade Copet, Devi Parikh, Yaniv Taigman, and Yossi Adi.
\newblock Audiogen: Textually guided audio generation.
\newblock \emph{arXiv preprint arXiv:2209.15352}, 2022.

\bibitem[Lee et~al.(2023)Lee, Jang, Chen, Qiu, and Huang]{post_prodct_2}
Yao-Chih Lee, Ji-Ze~Genevieve Jang, Yi-Ting Chen, Elizabeth Qiu, and Jia-Bin Huang.
\newblock Shape-aware text-driven layered video editing.
\newblock In \emph{Proceedings of the IEEE/CVF Conference on Computer Vision and Pattern Recognition}, pages 14317--14326, 2023.

\bibitem[Liu et~al.(2023)Liu, Chen, Yuan, Mei, Liu, Mandic, Wang, and Plumbley]{liu2023audioldm}
Haohe Liu, Zehua Chen, Yi Yuan, Xinhao Mei, Xubo Liu, Danilo Mandic, Wenwu Wang, and Mark~D Plumbley.
\newblock Audioldm: Text-to-audio generation with latent diffusion models.
\newblock \emph{arXiv preprint arXiv:2301.12503}, 2023.

\bibitem[Liu et~al.(2024{\natexlab{a}})Liu, Yuan, Liu, Mei, Kong, Tian, Wang, Wang, Wang, and Plumbley]{liu2024audioldm}
Haohe Liu, Yi Yuan, Xubo Liu, Xinhao Mei, Qiuqiang Kong, Qiao Tian, Yuping Wang, Wenwu Wang, Yuxuan Wang, and Mark~D Plumbley.
\newblock Audioldm 2: Learning holistic audio generation with self-supervised pretraining.
\newblock \emph{IEEE/ACM Transactions on Audio, Speech, and Language Processing}, 2024{\natexlab{a}}.

\bibitem[Liu et~al.(2024{\natexlab{b}})Liu, Zhang, Li, Lin, and Jia]{video_edit_2}
Shaoteng Liu, Yuechen Zhang, Wenbo Li, Zhe Lin, and Jiaya Jia.
\newblock Video-p2p: Video editing with cross-attention control.
\newblock In \emph{Proceedings of the IEEE/CVF Conference on Computer Vision and Pattern Recognition}, pages 8599--8608, 2024{\natexlab{b}}.

\bibitem[Luo et~al.(2024)Luo, Yan, Hu, and Zhao]{luo2024diff}
Simian Luo, Chuanhao Yan, Chenxu Hu, and Hang Zhao.
\newblock Diff-foley: Synchronized video-to-audio synthesis with latent diffusion models.
\newblock \emph{Advances in Neural Information Processing Systems}, 36, 2024.

\bibitem[OpenAI(2024)]{SoRA}
OpenAI.
\newblock Sora: Creating video from text.
\newblock \emph{https://openai.com/sora}, 2024.

\bibitem[Papamakarios et~al.(2021)Papamakarios, Nalisnick, Rezende, Mohamed, and Lakshminarayanan]{papamakarios2021normalizing}
George Papamakarios, Eric Nalisnick, Danilo~Jimenez Rezende, Shakir Mohamed, and Balaji Lakshminarayanan.
\newblock Normalizing flows for probabilistic modeling and inference.
\newblock \emph{Journal of Machine Learning Research}, 22\penalty0 (57):\penalty0 1--64, 2021.

\bibitem[PiKa(2024)]{PiKA}
PiKa.
\newblock Pika.
\newblock \emph{https://pika.art/try}, 2024.

\bibitem[Qi et~al.(2023)Qi, Cun, Zhang, Lei, Wang, Shan, and Chen]{post_prodct_1}
Chenyang Qi, Xiaodong Cun, Yong Zhang, Chenyang Lei, Xintao Wang, Ying Shan, and Qifeng Chen.
\newblock Fatezero: Fusing attentions for zero-shot text-based video editing.
\newblock In \emph{Proceedings of the IEEE/CVF International Conference on Computer Vision}, pages 15932--15942, 2023.

\bibitem[Radford et~al.(2021)Radford, Kim, Hallacy, Ramesh, Goh, Agarwal, Sastry, Askell, Mishkin, Clark, et~al.]{radford2021learning}
Alec Radford, Jong~Wook Kim, Chris Hallacy, Aditya Ramesh, Gabriel Goh, Sandhini Agarwal, Girish Sastry, Amanda Askell, Pamela Mishkin, Jack Clark, et~al.
\newblock Learning transferable visual models from natural language supervision.
\newblock In \emph{International conference on machine learning}, pages 8748--8763. PMLR, 2021.

\bibitem[Rai and Sridhar(2024)]{rai2024egosonics}
Aashish Rai and Srinath Sridhar.
\newblock Egosonics: Generating synchronized audio for silent egocentric videos.
\newblock \emph{arXiv preprint arXiv:2407.20592}, 2024.

\bibitem[Rombach et~al.(2022)Rombach, Blattmann, Lorenz, Esser, and Ommer]{rombach2022high}
Robin Rombach, Andreas Blattmann, Dominik Lorenz, Patrick Esser, and Bj{\"o}rn Ommer.
\newblock High-resolution image synthesis with latent diffusion models.
\newblock In \emph{Proceedings of the IEEE/CVF conference on computer vision and pattern recognition}, pages 10684--10695, 2022.

\bibitem[Sheffer and Adi(2023)]{sheffer2023hear}
Roy Sheffer and Yossi Adi.
\newblock I hear your true colors: Image guided audio generation.
\newblock In \emph{ICASSP 2023-2023 IEEE International Conference on Acoustics, Speech and Signal Processing (ICASSP)}, pages 1--5. IEEE, 2023.

\bibitem[Tian et~al.(2024)Tian, Wang, Zhang, and Bo]{tian2024emo}
Linrui Tian, Qi Wang, Bang Zhang, and Liefeng Bo.
\newblock Emo: Emote portrait alive-generating expressive portrait videos with audio2video diffusion model under weak conditions.
\newblock \emph{arXiv preprint arXiv:2402.17485}, 2024.

\bibitem[Van Den~Oord et~al.(2017)Van Den~Oord, Vinyals, et~al.]{van2017neural}
Aaron Van Den~Oord, Oriol Vinyals, et~al.
\newblock Neural discrete representation learning.
\newblock \emph{Advances in neural information processing systems}, 30, 2017.

\bibitem[Virtanen et~al.(2020)Virtanen, Gommers, Oliphant, Haberland, Reddy, Cournapeau, Burovski, Peterson, Weckesser, Bright, et~al.]{scipy}
Pauli Virtanen, Ralf Gommers, Travis~E Oliphant, Matt Haberland, Tyler Reddy, David Cournapeau, Evgeni Burovski, Pearu Peterson, Warren Weckesser, Jonathan Bright, et~al.
\newblock Scipy 1.0: fundamental algorithms for scientific computing in python.
\newblock \emph{Nature methods}, 17\penalty0 (3):\penalty0 261--272, 2020.

\bibitem[Wang et~al.(2024{\natexlab{a}})Wang, Ma, Pascual, Cartwright, and Cai]{wang2024v2a}
Heng Wang, Jianbo Ma, Santiago Pascual, Richard Cartwright, and Weidong Cai.
\newblock V2a-mapper: A lightweight solution for vision-to-audio generation by connecting foundation models.
\newblock In \emph{Proceedings of the AAAI Conference on Artificial Intelligence}, pages 15492--15501, 2024{\natexlab{a}}.

\bibitem[Wang et~al.(2024{\natexlab{b}})Wang, Wang, Wu, Song, Tan, Chen, Xu, and Sui]{wang2024tiva}
Xihua Wang, Yuyue Wang, Yihan Wu, Ruihua Song, Xu Tan, Zehua Chen, Hongteng Xu, and Guodong Sui.
\newblock Tiva: Time-aligned video-to-audio generation.
\newblock In \emph{Proceedings of the 32nd ACM International Conference on Multimedia}, pages 573--582, 2024{\natexlab{b}}.

\bibitem[Xing et~al.(2024)Xing, He, Tian, Wang, and Chen]{xing2024seeing}
Yazhou Xing, Yingqing He, Zeyue Tian, Xintao Wang, and Qifeng Chen.
\newblock Seeing and hearing: Open-domain visual-audio generation with diffusion latent aligners.
\newblock In \emph{Proceedings of the IEEE/CVF Conference on Computer Vision and Pattern Recognition}, pages 7151--7161, 2024.

\bibitem[Xu et~al.(2024{\natexlab{a}})Xu, Liu, Xing, Wang, Sun, Dan, Huang, Li, Cheng, Tai, et~al.]{TI_xuc}
Chao Xu, Yang Liu, Jiazheng Xing, Weida Wang, Mingze Sun, Jun Dan, Tianxin Huang, Siyuan Li, Zhi-Qi Cheng, Ying Tai, et~al.
\newblock Facechain-imagineid: Freely crafting high-fidelity diverse talking faces from disentangled audio.
\newblock In \emph{Proceedings of the IEEE/CVF Conference on Computer Vision and Pattern Recognition}, pages 1292--1302, 2024{\natexlab{a}}.

\bibitem[Xu et~al.(2024{\natexlab{b}})Xu, Li, Tu, Ren, Chen, Gu, Liang, and Yu]{xu2024video}
Manjie Xu, Chenxing Li, Xinyi Tu, Yong Ren, Rilin Chen, Yu Gu, Wei Liang, and Dong Yu.
\newblock Video-to-audio generation with hidden alignment.
\newblock \emph{arXiv preprint arXiv:2407.07464}, 2024{\natexlab{b}}.

\bibitem[Xue et~al.(2024)Xue, Deng, Gao, and Li]{xue2024auffusion}
Jinlong Xue, Yayue Deng, Yingming Gao, and Ya Li.
\newblock Auffusion: Leveraging the power of diffusion and large language models for text-to-audio generation.
\newblock \emph{arXiv preprint arXiv:2401.01044}, 2024.

\bibitem[Yang et~al.(2023)Yang, Yu, Wang, Wang, Weng, Zou, and Yu]{yang2023diffsound}
Dongchao Yang, Jianwei Yu, Helin Wang, Wen Wang, Chao Weng, Yuexian Zou, and Dong Yu.
\newblock Diffsound: Discrete diffusion model for text-to-sound generation.
\newblock \emph{IEEE/ACM Transactions on Audio, Speech, and Language Processing}, 31:\penalty0 1720--1733, 2023.

\bibitem[Zhang et~al.(2023)Zhang, Rao, and Agrawala]{zhang2023adding}
Lvmin Zhang, Anyi Rao, and Maneesh Agrawala.
\newblock Adding conditional control to text-to-image diffusion models.
\newblock In \emph{Proceedings of the IEEE/CVF International Conference on Computer Vision}, pages 3836--3847, 2023.

\bibitem[Zhang et~al.(2024{\natexlab{a}})Zhang, Mo, Zhang, and Morgado]{zhang2024audio}
Lin Zhang, Shentong Mo, Yijing Zhang, and Pedro Morgado.
\newblock Audio-synchronized visual animation.
\newblock \emph{arXiv preprint arXiv:2403.05659}, 2024{\natexlab{a}}.

\bibitem[Zhang et~al.(2024{\natexlab{b}})Zhang, Gu, Zeng, Xing, Wang, Wu, and Chen]{zhang2024foleycrafter}
Yiming Zhang, Yicheng Gu, Yanhong Zeng, Zhening Xing, Yuancheng Wang, Zhizheng Wu, and Kai Chen.
\newblock Foleycrafter: Bring silent videos to life with lifelike and synchronized sounds.
\newblock \emph{arXiv preprint arXiv:2407.01494}, 2024{\natexlab{b}}.

\end{thebibliography}
}


\end{document}